\documentclass [12pt]{article}
\usepackage[dvips]{graphicx}

\usepackage{epsfig}
\usepackage{graphicx}
\usepackage{indentfirst}
\usepackage{amsmath}
\usepackage{amsfonts}
\usepackage{amssymb}
\usepackage{hyperref}
 \usepackage{latexsym}
\usepackage{color}

\begin{document}
 \begin{center}
{\Large\bf Exploring stable models in $f(R, T, R_{\mu\nu}T^{\mu\nu})$ gravity.}\\

\medskip

 E. H. Baffou$^{(a)}$\footnote{e-mail:bahouhet@gmail},  
 M. J. S. Houndjo$^{(a,b)}$\footnote{e-mail:
sthoundjo@yahoo.fr} 
and J. Tossa$^{(a)}$\footnote{e-mail: joel.tossa@imsp-uac.org}

 $^a$ \,{\it Institut de Math\'{e}matiques et de Sciences Physiques (IMSP)}\\
  {\it 01 BP 613,  Porto-Novo, B\'{e}nin}\\
 $^{b}$\,{Universit\'e de Natitingou - B\'enin} \\
\end{center}
\begin{abstract}
We examine in this paper the stability analysis in $f(R,T, R_{\mu\nu }T^{\mu\nu})$ modified gravity , where $R$ and $T$ are the Ricci scalar
and the trace of the energy-momentum tensor, respectively. By considering the flat Friedmann universe, we obtain the \\ corresponding
generalized Friedmann equations and we evaluate the geometrical and matter perturbation functions.
The stability is developed using the de Sitter and power-law solutions. We search for application the stability of two particular cases of  $f(R,T, R_{\mu\nu }T^{\mu\nu})$
model  by solving numericaly the perturbation functions obtained.
\end{abstract}
 
\section{Introduction}

The study of the universe has undergone serious changes, as theoretical than observational during the last decades.
The cosmology, which attaches to understand the global properties and large structures of the universe (their origin, their evolution, theirs characteristics...)
is entered an era precision.
This discipline has long been a highly  speculative domain and reposed as much on metaphysics than physics.
With the formulation in $1915$ of the theory of General Relativity, a coherent theoretical framework allowing
mathematisation of space and time has become appeared.
One could then, in $1924$, formulate the cosmological models based on this theory.
These models of the universe, whose main characteristic is to be expanding, helped to understand many observations such 
as the recession of galaxies revealed by Edwin Hubble.
During twenty years, the cosmology is confined to the description and the reconstruction of this expansion.
In the second period, starting around $1948$, the study of physical processes in an expanding space allowed to formulate the big bang model.
This model based on Einstein's theory of $GR$ is one of the great succes stories in modern theoretical physic and it
describes a universe that is isotropic and homogeneous on large scales.
Recent observations of the supernovae type Ia (SNe Ia) \cite{A1}, the cosmic microwave background radiation (CMBR)
\cite{A2}, the baryon acoustic oscillation (BAO) surveys \cite{A3}, the large scale structure \cite{A4} and the weak lensing \cite{A5}, clearly
indicate that the universe is currently expanding with an accelerating rate.
These recent observations of the universe also possesses some problems for the standard model of cosmology and requires the presence
of a  \textquotedblleft dark energy \textquotedblright component whose nature is not well understood.
In order to explain the acceleration of the universe without introducing such a tantalizing source of energy, other gravitation theories have been designed.  
The Palatini $f(R)$ gravity is also proposed first as an alternative to dark energy in a series of  works \cite{D1}-\cite{D4}. There has since been growing interest
in these modified gravity theories: for the local tests of the Palatini and metric $f(R)$ gravity models see \cite{D5}-\cite{D6},
and for the cosmologies of these two classes of models see \cite{D7}-\cite{D13}.
A generalization of $f(R)$ theory, namely $f(R,T)$ gravity, was introduced first in Ref.\cite{Harko} and
then, studied in Refs.\cite{D14}-\cite{D25} with interesting results. The $f(R,T)$ gravity model depends on a source term, representing the variation
of the matter energy-momentum tensor with respect to the metric.
 The justification for the dependence on $T$
comes from inductions arising from some exotic fluid and/or quantum effects (conformal anomaly ). Actually, this
induction point of view adopts or links with the known proposals such as geometrical curvature inducing matter, a
geometrical description of physical forces, and a geometrical origin for the matter content of the universe.\par
Recently, $f (R, T)$ theory has been generalized by introducing the contribution from
contraction of the Ricci tensor and energy-momentum tensor named as $f(R,T, R_{\mu\nu} T^ {\mu\nu} )$
gravity \cite{Zahra}-\cite{Sergei}. In this theory, the energy-momentum tensor is generally non-conserved
 and extra force is present as a result of non-minimal matter geometry coupling. 
Haghani et al. \cite{Zahra} have discussed the cosmological implications of
this theory and developed the Dolgov-Kawasaki  instability criterion. The problem of matter instability in $f(R,T, R_{\mu\nu} T^{\mu\nu} )$
gravity is discussed  by Odintsov and S\`aez Gomez \cite{Sergei}.
Sharif and Zubair \cite{Sharif1}-\cite{Sharif2}
have presented the field equations for more general case and formulated energy conditions 
and the thermodynamic laws in $f(R,T, R_{\mu\nu} T^{\mu\nu} )$ modified gravity.
Because of the \\ interesting aspects and the progress realized by these authors,  it seems there is
still room to study some motivating gravitational and cosmological aspects of such modified gravity which have not yet been studied.
In the context of $f(R$) modified theories of gravity, the
stability of the Einstein static universe was also analyzed
by considering homogeneous perturbations \cite{D26}. By considering specific forms of $f(R)$, the stability regions of the
solutions were parameterized by a linear equation of state parameter $w = \frac{p}{\rho}$. In Ref. \cite{Baffou} the authors showed that the
cosmological viable $f(R,T)$ dark model obtained by imposing the conservation of the energy-momentum tensor
presents stability for both the de Sitter and power-law solutions.
In present paper, we extend the work of Ref. \cite{Baffou} in the context of $f(R,T, R_{\mu\nu} T^{\mu\nu} )$ modified gravity, ie 
we generalized the stability analysis in $f(R,T, R_{\mu\nu} T^{\mu\nu})$ gravity and analysed the viability for some specific known models.
This paper is outlined in the following manner: In Section $II$, we briefly review the basic formalism in 
$f( R,T, R_{\mu\nu }T^{\mu\nu})$ modified gravity. In section $III$, we consider a general form  of $f( R,T, R_{\mu\nu }T^{\mu\nu})$ gravity,
and analyze the stability solutions, by considering homogeneous perturbations around the Hubble parameter and energy density of matter.
Finally, in Section $IV$ we present our conclusions.

\section { Basic Formalism in $f( R,T, R_{\mu\nu }T^{\mu\nu})$  modified gravity }
In   $ f (R,T, R_{\mu\nu }T^{\mu\nu} ) $ gravity, we definie the action with matter as \cite{Zahra}-\cite{Sharif2}
\begin{eqnarray}
S = \frac{1}{2\kappa^{2}} \int \sqrt{-g} dx^{4} f \bigl(R,T,R_{\mu\nu}T^{\mu\nu}\bigr)+ \int \mathcal{L}_m \sqrt{-g}  dx^{4} \,, \label{1}
\end{eqnarray}
where $ R$, $T$ and  $ R_{\mu\nu }T^{\mu\nu}$ denote respectively the Ricci scalar, the trace of energy-momentum tensor 
and contraction of the Ricci tensor with $T^{\mu\nu}$.
The matter Lagrangian $\mathcal{L}_m$ is connected at energy-momentum tensor by the relation
 \begin{eqnarray}
T_{\mu\nu}=-\frac{2}{\sqrt{-g}}\frac{\delta\left(\sqrt{-g}\mathcal{L}_m\right)}{\delta g^{\mu\nu}}.\label{2}
\end{eqnarray} 
Assumption that the matter Lagrangian $\mathcal{L}_m$ is  a function of the metric $ g_{\mu\nu}$ and not its derivatives, the energy-momentum tensor (\ref{2}) takes the
following form
\begin{eqnarray}
T_{\mu\nu} = g_{\mu\nu}\mathcal{L}_{m}-\frac{{2}{\partial{\mathcal{L}_{m}}}}{\partial{g^{\mu\nu}}}. \label{3}
\end{eqnarray}
Varying the action (\ref{1}) with respect the components of the tensor metric $ g^{\mu\nu}$, one gets the field equations in $ f (R,T,R_{\mu\nu }T^{\mu\nu}) $ gravity given by 
\begin{eqnarray}
&& R_{\mu\nu}f_{R}-\Bigl[ \frac{1}{2}f-\mathcal{L}_{m}f_{T}- \frac{1}{2}\nabla_{\alpha}\nabla_{\beta}(f_{Q} T^{\alpha\beta})\Bigr] g_{\mu\nu}+
\bigl(g_{\mu\nu}\Box-\nabla_{\mu}\nabla_{\nu}\bigr) f_{R}+\frac{1}{2}\Box(f_{Q}T_{\mu\nu})+ \nonumber \\  
&& 2f_{Q}R_{\alpha}(\mu T^{\alpha}_\nu)-\nabla_{\alpha}\nabla_({\mu}\Big[ T^{\alpha}_{\nu)} f_{Q} \Big] -
G_{\mu\nu}\mathcal{L}_{m}f_{Q}-2\bigg(f_{T}g^{\alpha\beta}+f_{Q}R^{\alpha\beta}\bigg) 
\frac{\partial^{2}\mathcal{L}_{m}}{\partial{g^{\mu\nu}}\partial{g^{\alpha\beta}}}  \nonumber \\
&&= \bigg(k^{2}+f_{T}+\frac{1}{2}Rf_{Q}\bigg) T_{\mu\nu}, \label{4} 
\end{eqnarray}
where we setted $ Q = R_{\mu\nu}T^{\mu\nu}$ in order to simplify the expressions during this paper.\\ 
We note that when $ f (R,T, R_{\mu\nu }T^{\mu\nu} )\equiv f(R,T)$,  from Eq.(\ref{4}) we obtain the field equations in $f(R,T)$ gravity \cite{Harko} .
In the expression (\ref{4}) $f_{R}, f_{T}$ and $ f_{Q}$ represents the partial derivatives of $ f $ with respect to
 $ R $, $T$  and  $Q$ respectively; $G_{\mu\nu}$ being the Einstein tensor. \\
Contracting Eq.(\ref{4}) with respect the tensor metric $g^{\mu\nu}$, one gets the relation between the Ricci 
scalar $R$ and the trace $T$ of the energy-momentum tensor,
\begin{eqnarray}
&&(f_{R}+\mathcal{L}_{m}f_{Q})R+ \nabla_{\alpha}\nabla_{\beta}(f_{Q} T^{\alpha\beta})+4\mathcal{L}_{m}f_{T}-2f+3\Box f_{R}+  
\frac{1}{2}\Box(f_{Q}T)+ \nonumber \\
&& 2f_{Q}R_{\alpha\beta}T^{\alpha\beta}-2 g^{\mu\nu}\bigl(f_{T}g^{\alpha\beta}+f_{Q}R^{\alpha\beta}\bigr)\frac{\partial^{2}\mathcal{L}_{m}}{\partial{g^{\mu\nu}}\partial{g^{\alpha\beta}}}
= \bigg( k^{2}+f_{T}+\frac{1}{2}Rf_{Q}\bigg)T. \label{5}
\end{eqnarray}
By taking the divergence of the gravitational field equations (\ref{4}), we obtain the covariant divergence of the energy-momentum tensor as 
\begin{eqnarray}
&&\nabla_{\mu} T^{\mu\nu} = \frac{2}{(2+Rf_Q+2f_T)}\Bigg[\nabla_{\mu}(f_Q R^{\sigma \mu}T_{\sigma \nu})+\nabla_\nu (\mathcal{L}_m f_T)
-\frac{1}{2}(f_Q R_{\rho \sigma}+f_Tg_{\rho \sigma})\nabla_{\nu}T^{\rho\sigma}-\cr
&&G_{\mu\nu}\nabla^{\mu}(f_Q \mathcal{L}_m)-
\frac{1}{2}T_{\mu\nu} \biggl(\nabla^{\mu}(Rf_Q)+2\nabla^{\mu}f_T \biggr) \Bigg]. \label{5a}
\end{eqnarray}
Eq.(\ref{5a}) show that in  $ f (R,T, R_{\mu\nu }T^{\mu\nu} )$  theory the matter energy-momentum tensor is generally not conserved, and this non-conservation determines
the appearance of an extra-force acting on the particles in motion in the gravitational field. However in such theory we take often important aspect to guarante
the conservation of energy-momentum tensor.
We assume at follows the matter content of universe as perfect fluid which can be written as
\begin{eqnarray}
T_{\mu\nu} = ( \rho+ p )u_{\mu}u_{\nu} - p g_{\mu\nu}, \label{9}
\end{eqnarray}
where $\rho$ and $p$ are the energy density and the pressure of the
ordinary matter, respectively and the flat Friedmann-Robertson-walker (FRW) space time
with the element line 
\begin{eqnarray}
 ds^{2}= dt^{2}-a(t)^{2}\bigl[dx^{2}+ dy^{2}+dz^{2}\bigr].\label{10}
\end{eqnarray}
However in present study, we choose the Lagrangian density $ \mathcal {L}_{m} = -\rho $. 
We recall that the expression of the Lagrangian density is not unique, this expression depends on nature of matter source of the universe \cite{Harko}.
Therefore within the consideration that the Lagrangian density $ \mathcal {L}_{m} = -\rho $ does not depend on the metric tensor, 
 the  term $\frac{\partial^{2}\mathcal{L}_{m}}{\partial{g^{\mu\nu}}\partial{g^{\alpha\beta}}}$ vanish and the field equations (\ref{4}) becomes
\begin{eqnarray}
&& R_{\mu\nu}f_{R}-\Bigl[ \frac{1}{2}f+ \rho f_{T}- \frac{1}{2}\nabla_{\alpha}\nabla_{\beta}(f_{Q} T^{\alpha\beta})\Bigr] g_{\mu\nu}+
\bigl(g_{\mu\nu}\Box-\nabla_{\mu}\nabla_{\nu}\bigr) f_{R}+\frac{1}{2}\Box(f_{Q}T_{\mu\nu})+2f_{Q}R_{\alpha}(\mu T^{\alpha}_\nu)-\nonumber \\ 
&&\nabla_{\alpha}\nabla_({\mu}\Big[ T^{\alpha}_{\nu)} f_{Q} \Big] +
G_{\mu\nu} f_{Q}\rho  = \bigg(k^{2}+f_{T}+\frac{1}{2}Rf_{Q}\bigg) T_{\mu\nu}. \label{11} 
\end{eqnarray} 
To simplify, we interest  at the  general form of model  $ f(R,T,Q) = R + f(T) +f(Q)$, thus from the field equations (\ref{11}) the modified
Friedmann equations takes respectively the following forms
\begin{eqnarray}
 3H^{2}= \frac{1}{1+\rho f_{Q}}\Bigg[ (k^{2} +2 f_T)\rho+ \frac{1}{2}f(T,Q)+\frac{3}{2}H\partial_{t} \bigl[(p-\rho) f_{Q}\bigr]-
\frac{3}{2}\bigl(3H^2-\dot{H}\bigr)\rho f_Q - \frac{3}{2}\bigl(3H^2+\dot{H}\bigr)pf_Q \Bigg], \label{13}  
 \end{eqnarray}
 \begin{eqnarray}
&& -2\dot{H}-3H^{2}= \frac{1}{1+\rho f_{Q}}\Bigg[k^2 p+f_T\bigl(p-\rho\bigr)   -\frac{1}{2}f(T,Q)+\frac{1}{2}\bigl(\dot{H}+3H^2\bigr)\rho f_Q+ \cr
&& \frac{1}{2}\bigl(3H^2-\dot{H}\bigr)pf_Q  +  2H \partial_{t}\bigl[(p+\rho)f_Q\bigr]+\frac{1}{2}\partial_{tt}\bigl[(\rho-p)f_Q\bigr]  \Bigg], \label{14}
 \end{eqnarray}
where  we shall assume separable algebraic functions of the form $f(T,Q) = f(T)+f(Q)$.\par
When the content of matter is considered as ordinary matter with equation of state $p = w\rho$, we evaluate the expression $ Q=  R_{\mu\nu}T^{\mu\nu}$ as
\begin{eqnarray}
 Q = -3\rho \bigl[(1+w)\dot{H}+(1+3w)H^{2}\bigr]. \label{18}
\end{eqnarray}
Eqs.(\ref{13}) and (\ref{14}) can be reformulated in the following forms
\begin{eqnarray}
3H^{2}= \kappa^2 _{eff} \bigl (\rho + \rho_{DE}\bigr)=\kappa^2 _{eff} \rho_{eff} , \label{18a}  
\end{eqnarray}
\begin{eqnarray}
-2\dot{H}-3H^{2}= \kappa^2 _{eff} \bigl( p+p_{DE} \bigr)=\kappa^2 _{eff} p_{eff} ,  \label{18b}
\end{eqnarray}
where 
\begin{eqnarray}
\kappa^2 _{eff} = \frac{\kappa^2}{1+\rho f_Q}, 
\end{eqnarray}
\begin{eqnarray}
\rho_{DE}=  \frac{1}{\kappa^2}\Bigg[ 2 f_T\rho+ \frac{1}{2}f(T,Q)+\frac{3}{2}H\partial_{t} \bigl[(p-\rho) f_{Q}\bigr]-
\frac{3}{2}\bigl(3H^2-\dot{H}\bigr)\rho f_Q - \frac{3}{2}\bigl(3H^2+\dot{H}\bigr)pf_Q \Bigg],
\end{eqnarray}
\begin{eqnarray}
&& p_{DE} = \frac{1}{\kappa^2}\Bigg[ f_T\bigl(p-\rho\bigr)   -\frac{1}{2}f(T,Q)+\frac{1}{2}\bigl(\dot{H}+3H^2\bigr)\rho f_Q+ 
\frac{1}{2}\bigl(3H^2-\dot{H}\bigr)pf_Q  + \nonumber \\
&& 2H \partial_{t}\bigl[(p+\rho)f_Q\bigr]+\frac{1}{2}\partial_{tt}\bigl[(\rho-p)f_Q\bigr]  \Bigg],  
\end{eqnarray}
respectively.\par
From the conservation law, the effective energy density $\rho_{eff}$ evolves as
\begin{eqnarray}
\frac{d(\kappa^2 _{eff} \rho_{eff})}{dt}+3H\kappa^2 _{eff} (\rho_{eff}+p_{eff}) = 0.  \label{18cc}
\end{eqnarray}
Assuming the ordinary matter as dust and by making use of Eqs. (\ref{18a}) and (\ref{18b}), (\ref{18cc}) can be rewritten as
\begin{eqnarray}
f_{QQ} \dot{Q}+\bigl (3+2H-\frac{\dot{H}}{H}+\frac{\dot{\rho}}{\rho} \bigr)f_Q -\frac{1}{3H}\bigl(2\kappa^2+4f_T+\frac{1}{\rho}f(T,Q)-\frac{6}{\rho}H^2 \bigr) = 0. \label{18d}
\end{eqnarray}
One can rewrite the above equation using the redshift $z= \frac{1}{a} - 1$ as follows
\begin{eqnarray}
&&\frac{d^2 H}{dz^2}+\frac{1}{H} {\biggl(\frac{dH}{dz}\biggr)}^2+ \Bigg[1-\frac{5}{1+z}+\frac{f_Q}{3\rho_0 H^2 {(1+z)}^4}   \Bigg] \frac{dH}{dz}+ 
\frac{1}{9H^3 f_{QQ}}\Bigg[\frac{2(\kappa^2+2f_T)}{\rho_0{(1+z)}^5}+ \nonumber \\
&&\frac{1}{\rho_0^2{(1+z)}^8}\bigl(f(T,Q)-6H^2\bigr)\Bigg]+ 
\frac{(H-3)}{3\rho_0{(1+z)}^5}\frac{f_Q}{f_QQ}-\frac{3H}{{(1+z)}^2} =0 \label{18c}
\end{eqnarray}
Within the specific choice of $f(R,T,Q)$ model we can solve numericaly  Eq.(\ref{18c}) to analyze the evolution of cosmological parameters \cite{D23}
such as Hubble parameter $H(z)$, deceleration parameter $q(z)$, equation of state of dark energy $\omega_{DE}(z)$ and Ricci scalar $R(z)$   versus $z$.

\section{Stability analysis in f(R,T,Q) gravity}
In this section, we study the linear stability of the model $ f(R,T,Q) = R + f(T) +f(Q)$ using the power law and de
Sitter solutions. We will be interested to the perturbation of both the geometrical and matter parts of the modified equations of 
motion and we consider the ordinary matter as dust. For this, we focus our attention to the Hubble parameter for what concerns the geometry and the energy density
of the ordinary content (dust) concerning the matter of the background, and perform the perturbation about them as \cite{Felice}-\cite{Cruz}
\begin{eqnarray}
 H(t) = H_b(t) (1+\delta(t)), \quad  \rho(t)= \rho_{b}(t) (1+\delta_{m}(t)),
 \label{19}
\end{eqnarray}
where $H_b (t)$ and $\rho_b(t)$ satisfies the modified Friedmann equations (\ref{13})-(\ref{14}) and denote respectively the Hubble parameter and the
energy density of the ordinary matter of the background. $\delta(t)$ and $\delta_{m}(t)$ being, the perturbation functions about the geometry and the
matter, respectively.
Regarding the form of the $f(R,T,Q)$ model that we considered in this section, the novelty here is the effect coming from $f(T,Q) = f(T)+f(Q)$.
Then, we develop the model $f(T,Q)$ as function of two variable $T$ and $Q$ in a series of $ T_b = \rho_b$ and
$Q_b = -3\rho_b \bigl(\dot{H}_b +H_b^{2}\bigr)$ respectively, as follows
\begin{eqnarray}
f(T,Q)= f(T_b,Q_b) +\frac{\partial{f}}{\partial{T}}(T-T_b)+\frac{\partial{f}}{\partial{Q}}(Q-Q_b)+
\mathcal{O}^2, \label{20}
\end{eqnarray}
where $\mathcal{O}^2$ term includes all the terms proportional to the higher powers of $T$, $Q$ respectively.
From the standard conservation law of the energy-momentum tensor, we deduct the energy density of the ordinary matter at background as 
\begin{eqnarray}
\rho_{b}(t)=\rho_{0}\ e^{-3\int {H_{b}(t)}dt},\label{21}
\end{eqnarray}
with  $\rho_{0}$ the constant of integration. 
Considering that the ordinary matter is essentially the dust, the first modified Friedmann equations (\ref{13}) yields
\begin{eqnarray}
3H^{2}=\frac{1}{1+\rho f_{Q}} \Bigg[ (k^{2}+2f_T)\rho +\frac{1}{2}f(T,Q)-\frac{3}{2}H \partial_{t} (\rho f_{Q})-\frac{3}{2}(3H^2-\dot{H}) \rho f_{Q}\Bigg]. \label{22}  
\end{eqnarray}
By substituting the Eqs.(\ref{19}) and (\ref{20}) in (\ref{22}), one gets the following differential equation 
\begin{eqnarray}
&& \ddot{\delta}+\dot{\delta}\Bigg[ \frac{-2}{3\rho_b  H_b}f^{b}_{Q} +\biggl(4\rho_b H_b+\frac{\dot{\rho_b}}{\rho_b}+\frac{\dot{H_b}}{H_b} \biggr) f^{b}_{QQ}-3f^b_{QQQ}
\biggl(\dot{\rho_b}\dot{H_b}+\dot{\rho_b} H_b^2+2\rho_b H_b \dot{H_b} \biggr)\Bigg]+ \frac{H_b}{\rho_b}\dot{\delta_m}+  \cr
&& \delta \Bigg[ -\frac{4}{3\rho_b ^2}-\frac{2}{\rho_b}f^{b}_{Q}+f^{b}_{QQ}\biggl(-6+9H_b^2-\frac{\dot{H_b ^2}}{{H_b}^2}+2\frac{\dot{\rho_b}}{\rho_b}\frac{\dot{H_b}}{H_b}
+3\frac{\dot{\rho_b}}{\rho_b} H_b + 2 \frac{\ddot{H_b}}{H_b}+3\dot{H_b} \biggr)  \cr
&&-3f^{b}_{QQQ}\biggl( \frac{\dot{\rho_b}}{H_b} \dot{H_b^2}+3H_b\dot{H_b}\dot{\rho_{b}}
 +2\dot{\rho_b}{H_b}^3+\frac{\rho_b}{H_b}\dot{H_b}\ddot{H_b}+3\rho_b\ddot{H_b}+2\rho_b {\dot{H_b}}^2+4\rho_b H_b^2 \dot{H_b} 
\biggr) \Bigg ]+ \cr
&&\delta_m \Bigg[ \frac{2\kappa^2}{9\rho_b H_b ^2} -\frac{1}{\rho_b} f^b_{Q}+\frac{5}{9\rho_b H_b^2} f^b_{T}+\frac{1}{9H_b^2} f^b_{TT}+f^b_{QQ}
\biggl(2H_b^2-\frac{\dot{H_b ^2}}{H_b^2}+4\frac{\dot{\rho_b}}{\rho_b}\frac{\dot{H_b}}{H_b}+4\frac{\ddot{H_b}}{H_b}+4\frac{\dot{\rho_b}}{\rho_b} H_b+\frac{\dot{H_b}}{H_b}+3\dot{H_b} \biggr)\cr
&&-3f^{b}_{QQQ}\biggl(\frac{\dot{\rho_b}}{\rho_b}\dot{H_b}+3H_b\dot{H_b}\dot{\rho_b}+H_b^3\dot{\rho_b}+\rho_b\frac{\dot{H_b}}{H_b}\ddot{H_b}+\rho_b \dot{H_b}
+2\rho_b {\dot{H_b}}^2+2\rho_b \frac{\dot{H_b}}{H_b} \biggr)\Bigg]=0.
\label{23a}
\end{eqnarray}
Regarding the matter perturbation function, we obtain the differential equation
\begin{eqnarray}
\dot{\delta_{m}}(t)+3H_{b}(t)\delta(t) = 0. \label{24a}
\end{eqnarray}
By making use of Eqs.(\ref{23a}) and (\ref{24a}), we obtain the differential equation 
\newpage
\begin{eqnarray}
&& \dddot{\delta_m}+\ddot{\delta_m}\Bigg[-2\frac{\dot{H_b}}{H_b}-\frac{2}{3\rho_b H_b} f^{b}_{Q}+\biggl(4\rho_b H_b+  \frac{\dot{\rho_b}}{\rho_b}+ \frac{\dot{H_b}}{H_b}\biggr)f^{b}_{QQ} \nonumber \\      
&& -3f^{b}_{QQQ} \biggl(\dot{\rho_b}\dot{H_b}+H_b^2\dot{\rho_b}+2\rho_b H_b \dot{H_b}\biggr)\Bigg]  
 +\dot{\delta_m}\Bigg[-2\frac{\dot{H_b ^2}}{H_b^2}+\frac{\ddot{H_b}}{H_b}-3\frac{H_b^2}{\rho_b}-
 \frac{4}{3\rho_b^2}+ \nonumber \cr 
&& f^{b}_{Q}\bigl(\frac{2}{3}\frac{\dot{H_b}}{\rho_b}-\frac{2}{\rho_b}\bigr)+    
f^{b}_{QQ}\biggl(-6+9H_b^2-\frac{\dot{H_b}^2}{H_b^2}+2\frac{\dot{\rho_b}}{\rho_b}\frac{\dot{H_b}}{H_b}+3H_b\frac{\dot{\rho_b}}{\rho_b}+ 
2\frac{\ddot{H_b}}{H_b}+ 3\dot{H_b}-4\rho_b H_b^2\dot{H_b}- \nonumber \\
&& H_b\dot{H_b}\frac{\dot{\rho_b}}{\rho_b}-{\dot{H_b}}^2  \biggr)   
 -3f^{b}_{QQQ}\biggl(\frac{\dot{\rho_b}}{H_b}\dot{H_b}^2+
3H_b\dot{\rho_b}\dot{H_b}+2\dot{\rho_b}H_b^3-\frac{\rho_b}{H_b}\dot{H_b}\ddot{H_b}+3\rho_b\ddot{H_b}+2\rho_b\dot{H_b}^2 \cr
&& +  4\rho_bH_b^2\dot{H_b}- 
H_b\dot{H_b}^2 \dot{\rho_b}-H_b^3\dot{\rho_b}\dot{H_b}-2\rho_b H_b^2 \dot{H_b}^2 \biggr) \Bigg] +  
\delta_m \Bigg[ \frac{2\kappa^2}{9\rho_b H_b ^2} -\frac{1}{\rho_b} f^b_{Q}+\frac{5}{9\rho_b H_b^2} f^b_{T}+\frac{1}{9H_b^2} f^b_{TT}+\cr  
&& f^b_{QQ} \biggl(2H_b^2-\frac{\dot{H_b ^2}}{H_b^2}+4\frac{\dot{\rho_b}}{\rho_b}\frac{\dot{H_b}}{H_b}+4\frac{\ddot{H_b}}{H_b}+4\frac{\dot{\rho_b}}{\rho_b} H_b+\frac{\dot{H_b}}{H_b}+3\dot{H_b} \biggr)
-3f^{b}_{QQQ}\biggl(\frac{\dot{\rho_b}}{\rho_b}\dot{H_b}+3H_b\dot{H_b}\dot{\rho_b}+H_b^3\dot{\rho_b}+ \cr
&&\rho_b\frac{\dot{H_b}}{H_b}\ddot{H_b}+\rho_b \dot{H_b} +2\rho_b {\dot{H_b}}^2+2\rho_b \frac{\dot{H_b}}{H_b} \biggr)\Bigg]=0.
\label{23}
 \end{eqnarray}
 Once the numerical solution of the equation (\ref{23}) be done, we can perform the evolution of the geometrical perturbation function $\delta(t)$ through the equation (\ref{24a}).
 To illustrate the viability (stability) of the $f( R,T, Q)$ model, we interest in next subsection at the evolution of geometrical and matter perturbation functions 
$\delta$  and $\delta_m$ within the de Sitter and power-law solutions. To do, we consider two specific actions \cite{Zahra}
\begin{eqnarray}
   1. \quad   f(R,T,Q) = R +\alpha Q, \\
   2. \quad   f(R,T,Q) = R +\alpha Q + \beta \sqrt{T},
\end{eqnarray}
where $\alpha$ and $\beta$ are constant parameters.
These models have been studied in \cite{Zahra} to analyze the evolution and dynamics of the universe for the above with and without energy conservation.
Recently, the energy conditions and the thermodynamic laws of these particular models have been investigated by the authors of the references 
\cite{Sharif1}-\cite{Sharif2}.
 \subsection{ Stability of de Sitter solutions}
 In de Sitter solutions, the Hubble parameter for the background is a constant and one has $ H_b (t) = H_0$.
 Thus the equations (\ref{21}) and (\ref{23}) reduces to
 \begin{eqnarray}
\rho_{b}(t)=\rho_{0}\ e^{-3H_{0}t},\label{24}
\end{eqnarray}
\begin{eqnarray}
 \dddot{\delta_m}+c_1 \ddot{\delta_m}+c_2\dot{\delta_m}+c_3\delta_m=0, \label{25}
\end{eqnarray}
where
\begin{eqnarray}
c_1=-\frac{2}{3\rho_b H_0}f^{b}_{Q}+H_0f^{b}_{QQ}\bigl(4\rho_b-3 \bigl)-3H_0^2\dot{\rho_b}f^{b}_{QQQ}, \label{26} 
\end{eqnarray}
\begin{eqnarray}
c_2= -\frac{3H_0^2}{\rho_b}-\frac{4}{3\rho_b^2}-\frac{2}{\rho_b}f^{b}_{Q} -6f^{b}_{QQ}-6H_0^3\dot{\rho_b}f^{b}_{QQQ}, \label{27} 
\end{eqnarray}
\begin{eqnarray}
c_3= \frac{2\kappa^2}{9\rho_b H_0^2}-\frac{1}{\rho_b}f^{b}_{Q}+\frac{5}{9\rho_bH_0^2}f^{b}_{T}+\frac{1}{9H_0^2}f^{b}_{TT}-10H_0^2 f^{b}_{QQ}-3H_0^3\dot{\rho_b}f^{b}_{QQQ}.\label{28}
\end{eqnarray}
\begin{figure}[h]
\centering
\begin{tabular}{rl}
\includegraphics[width=8cm, height=8cm]{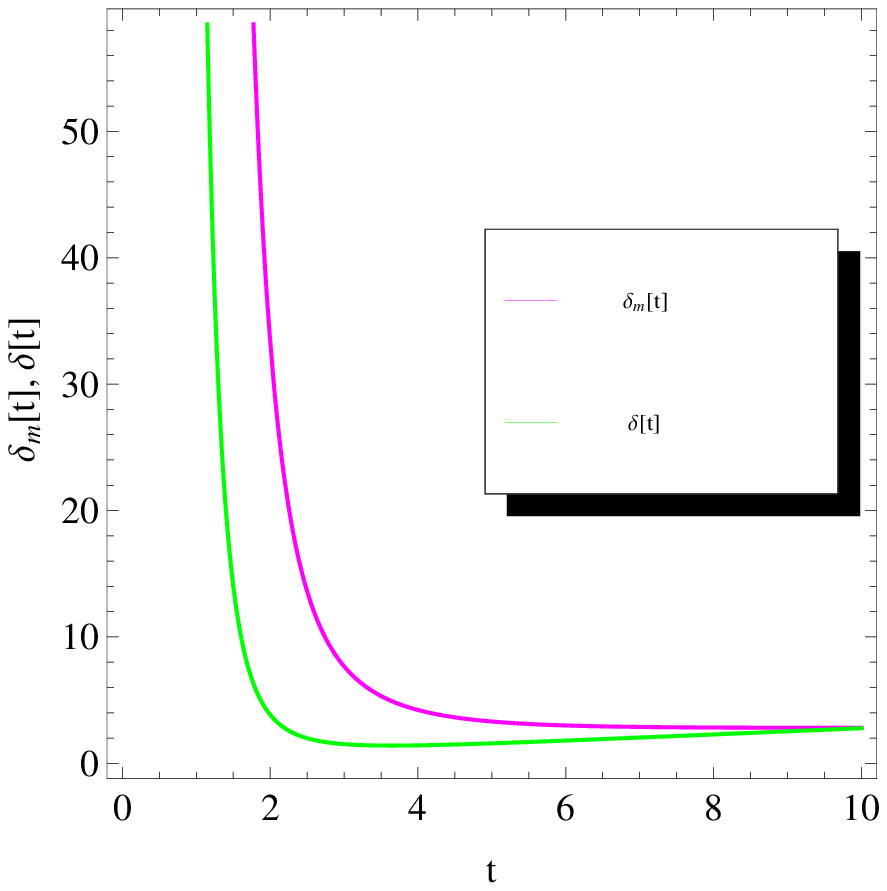}&
\includegraphics[width=8cm, height=8cm]{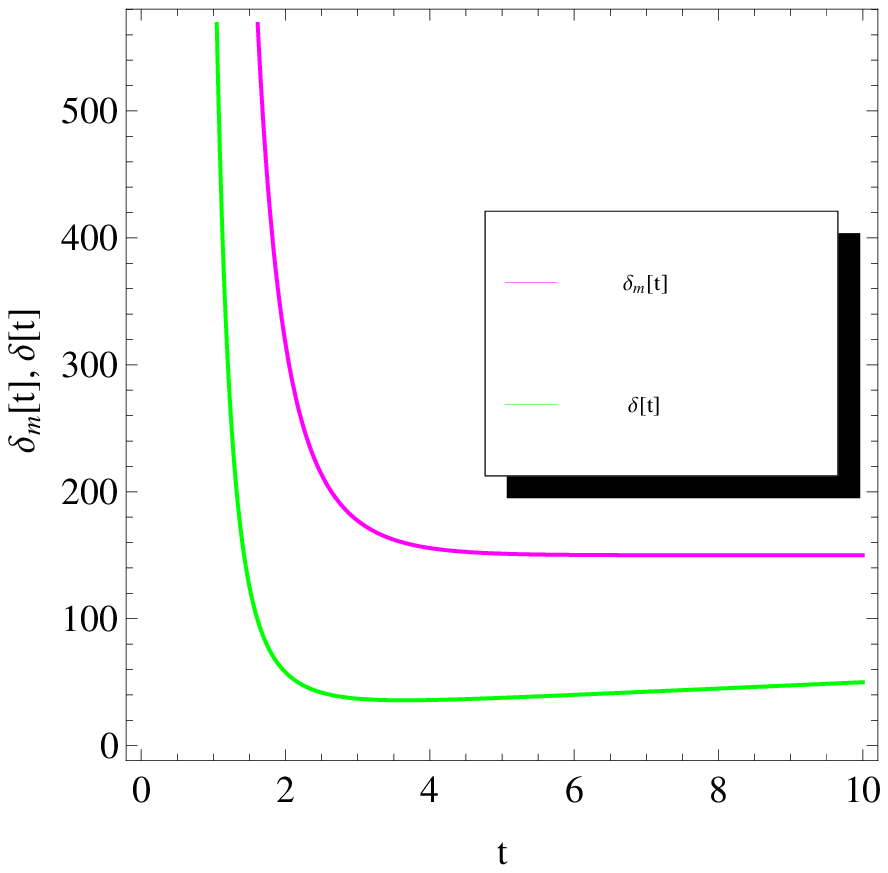}
\end{tabular}
\caption{The figures show the evolution of the perturbation functions $\delta_m$ and $\delta $  within the de Sitter solutions.
The graph at the left side of the figure \ref{fig1} represents the evolution of these perturbation functions for the model $f(R,T,Q)= R +\alpha Q$, while
the one at the right side show the evolution of the perturbation functions for the model $f(R,T,Q) = R +\alpha Q + \beta \sqrt{T}$. 
The graphs are plotted for $\rho_0 =1$, $\alpha=3$, $\kappa =1.5$,  $ H_0 = 0.3$ and $\beta = 1.5$.}
\label{fig1}
\end{figure}
\subsection{ Stability of power-law solutions}
In this fact, we are interested to the cosmological solutions of the form $ a(t)\propto t^{n}\Rightarrow H_b(t) = \frac{n}{t}.$
For this, the equations (\ref{21}) and (\ref{23}) are becomes  
\begin{eqnarray}
\rho_b(t) = \rho_0 t^{-3n}   \label{29}
\end{eqnarray}
\begin{eqnarray}
 \dddot{\delta_m}+ c'_1 \ddot{\delta_m}+c'_2\dot{\delta_m}+c'_3\delta_m=0, \label{30}
\end{eqnarray}
where
\begin{eqnarray}
c'_1 =\frac{2}{t}-\frac{2}{3\rho_0 n} t^{3n+1}f^{b}_{Q}+ \bigg (\frac{4\rho_0 n}{t^{{3n+1}}}-\frac{(3n+1)}{t} \bigg) f^{b}_{QQ}+
-\frac{3n^2 \rho_0}{t^{3n+3}}(1-n)f^{b}_{QQQ},\label{31} 
\end{eqnarray}
\newpage
\begin{eqnarray}
&& c'_2 = \frac{-3n^2}{\rho_0 t^{2-3n}}-\frac{4}{3\rho_0^2} t^{6n}-\frac{2}{\rho_0}\bigg(\frac{n}{3t^{2-3n}}+t^{3n}\bigg)f^{b}_{Q}+f^{b}_{QQ}
\bigg(-6-\frac{1}{t^2}(5-9n)+\frac{1}{t^4}(n^2-3n^3)+\frac{4n^2 \rho_0}{t^{3n+4}} \bigg) \cr
&&-3\rho_0 n f^{b}_{QQQ}\bigg(-\frac{6n^3}{t}+9n^2 t^{3n-4}+
\frac{1}{t^{3n+4}}(3n^4-n+2)+\frac{6}{t^{3n+3}}+\frac{5n^3}{t^{3n+6}}\bigg), \label{32} 
\end{eqnarray}
\begin{eqnarray}
&& c'_3 = \frac{1}{9\rho_0 n^2}(2\kappa^2+5f^{b}_{T})t^{3n+2}-\frac{t^{3n}}{\rho_0}f^{b}_{Q}  +\frac{t^2}{9n^2}f^{b}_{TT}+\bigg(-\frac{1}{t}-\frac{1}{t^2}(10n^2-9n+1)+\frac{8n^2}{t^4}\bigg)f^{b}_{QQ} \cr
&&-3f^{b}_{QQQ}\bigg(\frac{3n^2}{t^3}+\frac{\rho_0}{t^{3n+4}}(9n^3+2n^2-2n)-\frac{3n^4 \rho_0}{t^{3n-2}}-\frac{n \rho_0}{t^{3n+2}}-\frac{2\rho_0}{t^{3n+1)}}\bigg).\label{32}             
\end{eqnarray}
 
\begin{figure}[h]
\centering
\begin{tabular}{rl}
\includegraphics[width=8cm,height=8cm]{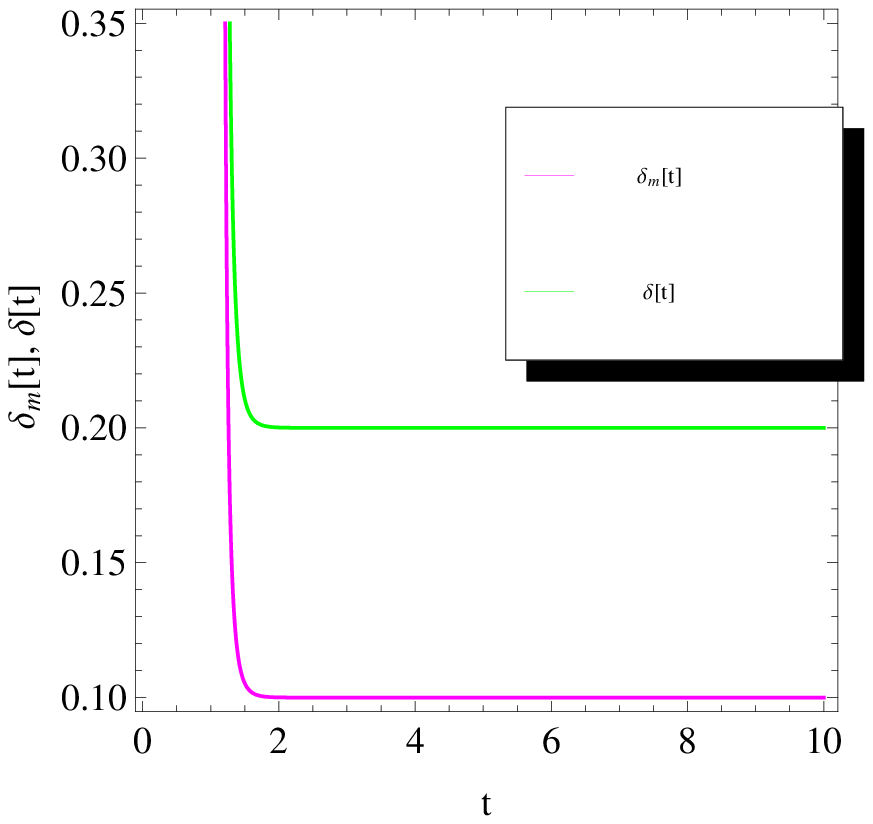}&
\includegraphics[width=8cm,height=8cm]{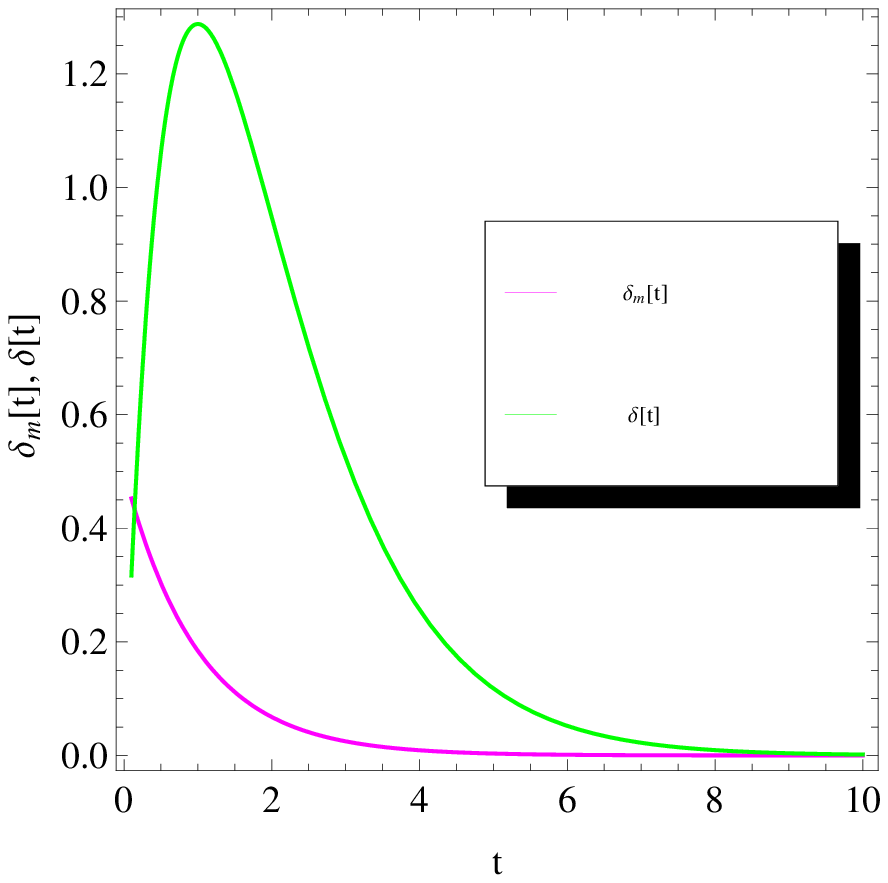}
\end{tabular}
\caption{The figures show the evolution of the perturbation functions $\delta_m$ and $\delta $  within the power-law solutions.
The graph at the left side of the figure \ref{fig2} represents the evolution of these perturbation functions for the model $f(R,T,Q)= R +\alpha Q$, while
the one at the right side shows the evolution of the perturbation functions for the model $f(R,T,Q) = R +\alpha Q + \beta \sqrt{T}$. 
The graphs are plotted for $\rho_0 =1$, $\alpha =0.1$, $\kappa =1.5$,  $ H_0 = 2$, $n=0.5$ and $\beta = 1.5$.  }
\label{fig2}
\end{figure}
Regarding the evolution of the perturbation functions plotted in figures \ref{fig1} and \ref{fig2}, we see that when the universe expands i.e the
time evolves, the matter and geometric perturbation functions $ \delta_m$ and $\delta$ converge, respectively. 
We conclude that the both models considered presents stability through the convergence of the geometrical and \\ matter perturbation 
functions $ \delta_m$ and $\delta$ within the de Sitter and power-law solutions .

\newpage
\section{Conclusion} 
In this present paper, we have performed the stability analysis in gravitational theory in which
matter is coupled to geometry, with
the effective Lagrangian of the gravitational field being given by an arbitrary function of the Ricci
scalar $R$, the trace of the matter energy-momentum tensor $T$, and the contraction of the Ricci tensor with
the matter energy-momentum tensor $Q= R_{\mu\nu} T^{\mu\nu}$. By chosing the general form $f(R,T,Q)= R+f(T)+f(Q)$, we obtained the modified FRW field
equations. In order to check the viability of this model, we establish the differential equation of matter and geometric perturbation functions and we perform
its stability taking into account the de Sitter and power-law cosmological solutions.
To illustrate how these perturbation functions ($ \delta_m$ and $\delta$) can constrain the $ f(R,T,Q)$ gravity, we have taken two particular cases, namely
 $f(R,T,Q) = R +\alpha Q$ and $f(R,T,Q) = R +\alpha Q + \beta \sqrt{T}$. We see that for the both considered solutions, the models considered presents stability through the convergence of the 
geometric and  matter  perturbation functions $\delta$ and $\delta_m$. The higher derivatives $f(R,T,Q)$ gravity is very complex
and can open a new perspective on the very early stages of the evolution of the Universe.


\vspace{1.5cm}

{\bf Acknowledgement}: The authors thank Prof. S. D. Odintsov for useful comments and suggestions.
E. H. Baffou  thanks IMSP for every kind of support during the realization of this work.
M. J. S. Houndjo  thanks {\it Universit\'e de Natitingou}.


\end{document}